\documentclass[useAMS,usenatbib]{mn2e}
\usepackage[dvips]{graphicx}
\usepackage{subfigure}

\title[Frequency, splitting, linewidth and amplitude estimates of
low-$\ell$ p modes of $\alpha$~Cen~A]{Frequency, splitting,
linewidth and amplitude estimates of low-$\ell$ p modes of
$\alpha$~Cen~A: analysis of WIRE photometry}
\author[S. T. Fletcher et al.]
{S. T. Fletcher$^{1}$\thanks{E-mail:stfletch@bison.ph.bham.ac.uk},
W. J. Chaplin$^{1}$, Y. Elsworth$^{1}$, J. Schou$^{2}$ and D. Buzasi$^{3}$\\
$^{1}$School of Physics and Astronomy, University of Birmingham,
Edgbaston, Birmingham, B15 2TT, UK\\
$^{2}$Stanford University, HEPL Annex 201, Stanford, CA 94305-4085, USA\\
$^{3}$Department of Physics, US Air Force Academy, Colorado Springs, CO 80840, USA}

\begin{document}

\maketitle

\begin{abstract}
We present results of fitting the 50-day time series of photometry of $\alpha$ Cen A taken by the WIRE satellite
in 1999. Both power spectrum and autocovariance function (ACF) fitting techniques were used in an attempt to
determine mode frequencies, rotational splittings, lifetimes and amplitudes of low-$\ell$ p-modes. In all, using
both techniques, we managed to fit 18 modes (seven $\ell$ = 0, eight $\ell$ = 1 and three $\ell$ = 2) with
frequencies determined to within 1 - 2 $\mu$Hz. These estimates are shown to be 0.6 $\pm$ 0.3 $\mu$Hz lower, on
average, than the frequencies determined from two other more recent studies (\citeauthor{Bouchy2002}
\citeyear{Bouchy2002}; \citeauthor{Bedding2004} \citeyear{Bedding2004}), which used data gathered about 19
months after the WIRE observations. This could be indicative of an activity cycle, although due to the large
uncertainty, more data would be needed to confirm this.

Over a range of 1700 to 2650 $\mu$Hz we were also able to use the ACF fitting to determine an average lifetime
of 3.9 $\pm$ 1.4 days, and an average rotational splitting of 0.54 $\pm$ 0.22 $\mu$Hz, which is the first ever
reliable estimate of this parameter. In contrast to the ACF, the power spectrum fitting was shown to return
significantly biased results for these parameters.
\end{abstract}

\begin{keywords}
stars: oscillations -- methods: data analysis
\end{keywords}

\section{Introduction}\label{Intro_Sec}
The past ten years have seen a number of increasingly successful attempts to detect and measure solar-like
oscillations in other stars. Due to its proximity and similarity to the Sun, many of these studies have been
focused on the star $\alpha$ Cen A. The first clear detection of p-mode oscillations on this star was made by
\cite{Schou2001} from photometry using the Wide-Field Infrared Explorer (WIRE) satellite taken over a 50-day
period. \cite{Schou2001} correctly determined the large frequency separation but, unfortunately, wrong $\ell$
identifications were made and hence an incorrect value for the small separation was determined. Further
detections and the first correct mode identifications were made by \cite{Bouchy2002} using a 13-day run of
velocity measurements taken by the CORALIE spectrograph. More recently, \cite{Bedding2004} determined the
frequencies for over 40 individual modes from observations by the UVES and UCLES instruments taken over a period
of 5 nights.

The main driving force behind each subsequent study has been to improve the signal-to-noise ratio (SNR) in order
to initially detect as many modes as possible and then to better constrain the limits placed on the determined
frequencies. It is of course also important to improve resolution, but practical constraints have meant all
these studies were limited in the length of observations that could be made. This has meant accurate
determination of mode parameters such as power, rotational splitting and lifetime has been difficult.

Here, we apply two sophisticated fitting procedures to the WIRE $\alpha$ Cen A data collected in 1999 in order
to improve the parameter determinations. Although this data set has the poorest SNR of the three studies
mentioned above, it does have the longest time series. Hence, we would expect to extract more reliable estimates
of the average lifetime and rotational splitting of the $\alpha$ Cen A modes.

The first fitting procedure we applied was a traditional power spectrum fitting method. This involved taking the
Fourier transform of the time series and then fitting a Lorentzian-like model to the various mode peaks in the
resulting power spectrum. The second procedure used is a new technique based on fitting the autocovariance
function (ACF) of the time series (i.e., the unnormalized autocorrelation function). Since the modes seen in the
ACF are all superimposed, one must first filter the time series in order to isolate the modes one is hoping to
fit. The ACF can then be computed and a model based on an exponentially decaying, periodic function fitted to
the result. This technique was first introduced as a possible method for mode determination by
\cite{Gabriel1998} and developed more fully by \cite{Fletcher2004} in an attempt to better constrain mode
parameters of long solar p-mode data sets.

In the initial investigation of the WIRE data by \cite{Schou2001} only a handful of modes were identified due to
the poor SNR in the data set. However, a distinct advantage of revisiting this data comes in having a large
number of robust mode identifications from the aforementioned two later studies. This provides additional
a-priori information that we can use as initial `guess' values for our fitting procedures.

In order to fully test our fitting of the WIRE data we also generated a set of artificial time series. These
data were created specifically to mimic the WIRE data allowing us to explicitly determine the precision and
robustness of our fitted parameters. We detail the creation of this simulated data in Section~\ref{Data_Sec}. In
Section~\ref{ModelFit_Sec} we describe in detail the procedure involved in applying the two fitting techniques
to the data. Finally, in Section~\ref{Results_Sec}, we go on to present and analyse the results of our fitting
for the mode frequencies, amplitudes, lifetimes and rotational splitting parameters.

\section{Data}\label{Data_Sec}

\begin{figure}
\centering \subfigure[50 days] {\label{Wite_TimeSeries_50d}\includegraphics[width=3.3in]{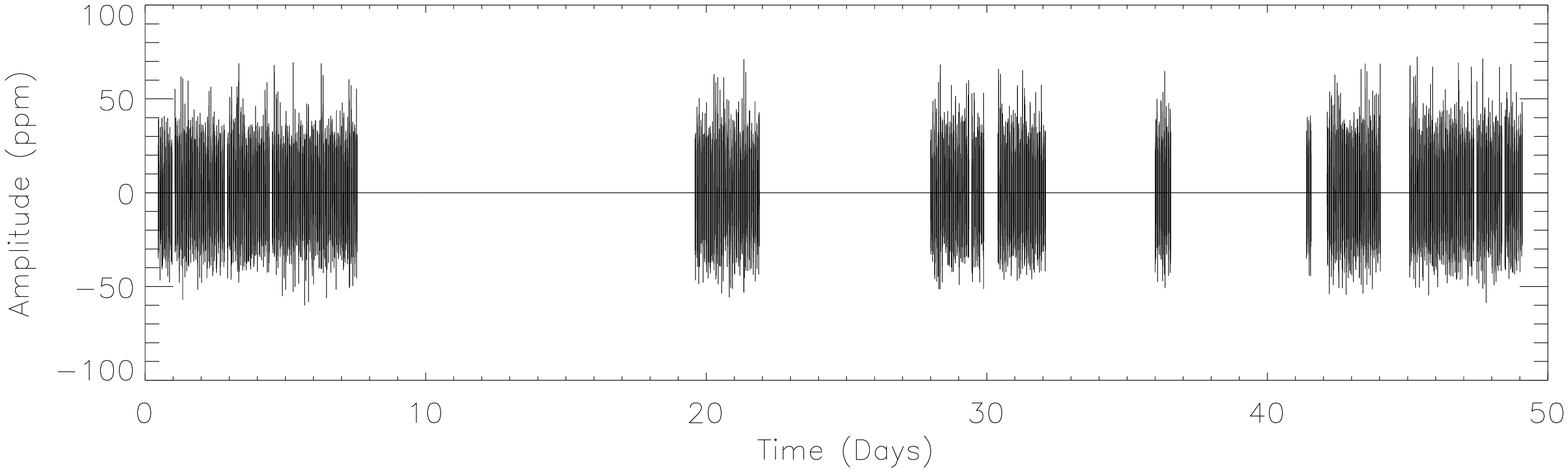}}
\subfigure[8 days] {\label{Wite_TimeSeries_8d}\includegraphics[width=3.3in]{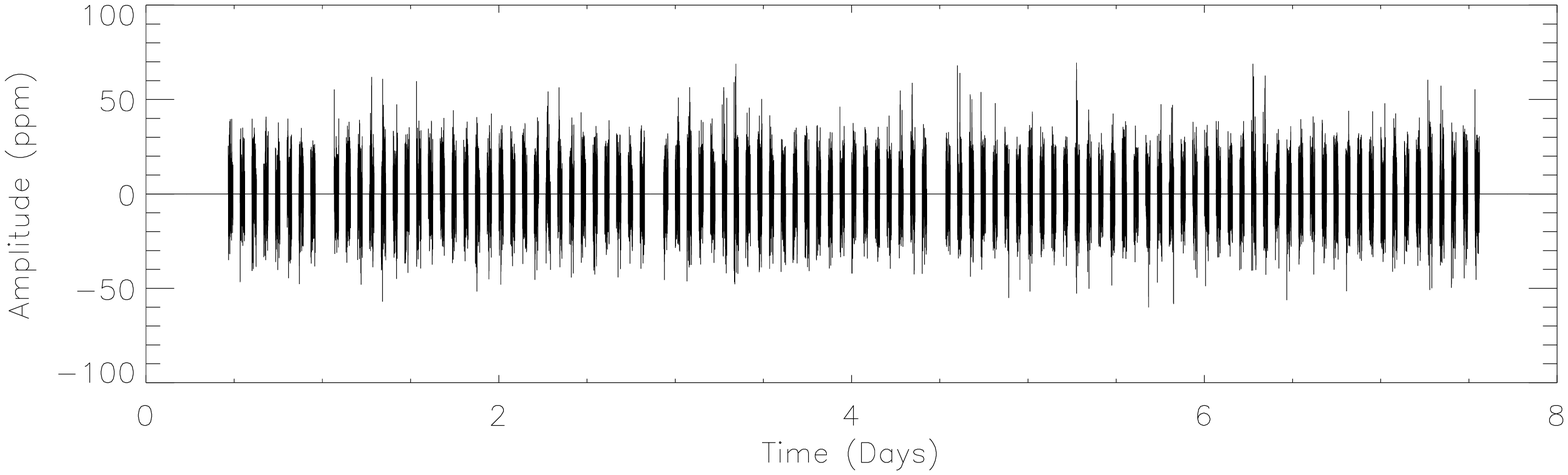}} \caption{The
Wire $\alpha$ Cen A time series over (a) the full 50 days and (b) the first 8 days only.}
\label{Wire_TimeSeries}
\end{figure}

The WIRE satellite collected 50 days of photometry observations on the star $\alpha$ Cen A between July 15 and
September 3, 1999. The orbit was 96 minutes, low Earth and Sun-synchronous, which unfortunately lead to rather
significant pointing constraints. Hence, observations could only be made for about 40 minutes out of each orbit.
Additionally there were some days when no data were collected and some days where data could not be used due to
processing problems. Hence, the resulting duty cycle for the WIRE time series was only 15\%. The processing of
the WIRE data to produce the final time series is described in \cite{Buzasi2000} and \cite{Schou2001}. The data
were binned to a 10s cadence resulting in a series of 432,000 points. The processed time series is shown in
Fig~\ref{Wire_TimeSeries}, for both the full 50 days and a zoomed in section with just the first 8 days in order
to show the main gap structure. Being as the time series was made using photometry observations, the amplitudes
are given in parts-per-million (ppm). The method for converting between ppm and velocity is described in
Section~\ref{Amps_Sec}.

A full set of simulated data made to resemble the WIRE time series was produced in order to test fitting on the
real data. The Laplace transform solution of the equation of a forced, damped harmonic oscillator was used in
order to generate individual mode components of the artificial time series. The application of this model is
described more fully in \cite{Chaplin1997}. A set of low-$\ell$ modes covering the ranges 0 $ \leq \ell \leq $ 2
and 1700 $ \leq \nu \leq $ 2650 $\mu$Hz was created in this way. It should be noted that the $\ell$ = 3 modes
are not included since they do not appear to be detectable in the WIRE photometry data, although for more recent
velocity observations they are \citep{Bedding2004}. For each $\ell \neq 0$, a rotationally induced splitting
pattern was included, for which the visibilities, $\varepsilon$, of each component were fixed using the
following equations (\citeauthor{Gizon2003} \citeyear{Gizon2003}; \citeauthor{Toutain1993}
\citeyear{Toutain1993})
\begin{equation}
\varepsilon_{1,0} = \cos^2 i,\label{VisInc1}
\end{equation}
\begin{equation}
\varepsilon_{1,\pm1} = {\textstyle\frac{1}{2}} \sin^2
i,\label{VisInc2}
\end{equation}
\begin{equation}
\varepsilon_{2,0} = {\textstyle\frac{1}{4}} (3 \cos^2 i
-1)^2,\label{VisInc3}
\end{equation}
\begin{equation}
\varepsilon_{2,\pm1} = {\textstyle\frac{3}{8}} \sin^2
(2i),\label{VisInc4}
\end{equation}
\begin{equation}
\varepsilon_{2,\pm2} = {\textstyle\frac{3}{8}} \sin^4
i,\label{VisInc5}
\end{equation}
and the inclination, $i$, was chosen as
$79^\textup{\footnotesize{o}}$ \citep{Pourbaix1999}.

A database of frequency and powers, based on preliminary fits to the WIRE data, was used in order to fix the
various characteristics of each mode. Where no modes could be identified, frequencies presented in
\cite{Bedding2004} were used.

Based on fits to the ACF of the WIRE data (see section~\ref{ACF_Sec}) we chose a constant rotational splitting
of 0.5 $\mu$Hz and a constant linewidth (full width half maximum) of 1.0 $\mu$Hz (equivalent to a lifetime or
e-folding amplitude time of $\sim$3.7 days). Analysis of fits to these artificial data (see sections 4.2 and
4.3) showed the parameters could be robustly determined. This gives confidence that the simulated spectra were a
good match to, and a reliable cross check for, the real data. A background offset (white noise) was also added
so as to match the WIRE data. In total 50 independent data sets were created, each of a length equal to the WIRE
time series. The WIRE window function was imposed on each of them.

\section{Modelling and Fitting}\label{ModelFit_Sec}

In this section we outline the two different methods used to determine the various p mode parameters. First we
discuss the more traditional power spectrum technique and then go onto introduce the ACF method. In both cases
we explain the specific models used to fit the WIRE data.

\subsection{Power Spectrum}\label{PowSpec_Sec}

When fitting the power spectrum of a solar or solar-like p-mode time series the various mode components are
generally modelled using either a Lorentzian, or an asymmetric function that models small departures from a
symmetric shape \citep{Nigam1998}. For low-$\ell$ (Sun-as-a-star) data this is commonly done by fitting
individual mode pairs (i.e., $\ell$ = 0 with neighboring $\ell$ = 2; and $\ell$ = 1 with $\ell$ = 3) using
narrow fitting `windows' centered on the target pair.

\begin{figure}
\centering \subfigure[The WIRE Power Spectrum]
{\label{WirePowerSpec}\includegraphics[width=3.3in]{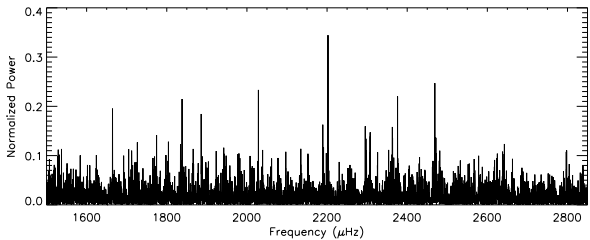}} \subfigure[The WIRE Spectral Window]
{\label{WireWinPowerSpec}\includegraphics[width=3.3in]{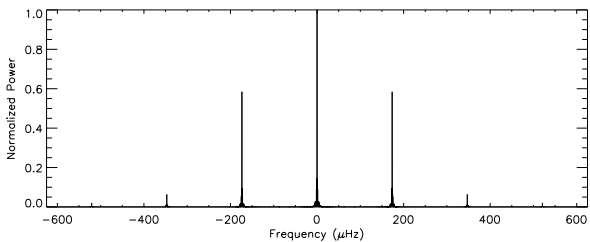}} \caption{A (a) The WIRE Power Spectrum
over the region 1500 $ \leq \nu \leq $ 2850 $\mu$Hz and (b) the spectral window plotted over the same total
frequency range.} \label{PowerSpec}
\end{figure}

Unfortunately, certain characteristics of the WIRE data make this strategy difficult to implement. The rather
severe window function mentioned in Section~\ref{Data_Sec} results in prominent sidebands at 173.6 $\mu$Hz from
the mode peaks. As a result the sidebands of modes lie in the vicinity of other overtones. This makes it very
difficult to isolate a single mode pair and its sidebands without the presence of intervening modes. Therefore,
we adopt a strategy of fitting Lorentzian profiles to all modes simultaneously. Note that the WIRE power
spectrum and the spectral window are shown in Fig.~\ref{PowerSpec}

In the 1700 $ \leq \nu \leq $ 2650 $\mu$Hz range a total of 27 modes were included in the simulated data sets: 9
each of $\ell$ = 0, 1 and 2. The $\ell$ = 1 and 2 modes were split into three and five components respectively
to match the rotational splitting pattern. However, we chose to fit only the modes that are identifiable in the
WIRE data, of which there are eighteen.

For the power spectrum method, the power spectral density, $P$, was modelled using a standard Lorentzian profile
for each peak summed over all visible modes and corresponding sidebands in the chosen frequency range, offset
with a flat background, $B$, i.e.,
\begin{equation}
P = B+\sum_{nlmk} H_{nlmk} Z_{nlmk}^{-1},
\end{equation}
where
\begin{displaymath}
Z_{nlmk} = 1+ \left( \frac{ \nu - \nu_{nl} + ms + k\textup{w} }
{\Delta/2} \right)^2.
\end{displaymath}
Here, $H$ is the height of each mode component (i.e., maximum
power spectral density), $\nu_{nl}$ is the central frequency of a
mode, $s$ is the rotational splitting, w is the sideband spacing
and $\Delta$ is the width.

In the WIRE data, over the range 1700 $ \leq \nu \leq $ 2650 $\mu$Hz, the radial overtone number, $n$, increases
from 14 to 23. However, since we fit only 18 modes we do not sum over all the possible $n\ell$ combinations.
Table 1 gives the list of $n$ and $\ell$-valued modes that are fitted. $m$ is the azimuthal order and is summed
over $-\ell \leq m \leq \ell$, while $k$ is a dummy variable allowing us to include the first order sidebands in
our model and as such is summed over $-1 \leq k \leq 1$.

\begin{table}
\caption{$n$ and $\ell$ values included in fitting.} \centering
\begin{tabular}{ccccc}
\hline $n$&$\ell$&$n$&$\ell$\\ \hline
14&2    &19&0,1  \\
15&1    &20&0,1  \\
16&0,1  &21&0,1,2\\
17&0,1,2&22&0    \\
18&1    &23&0,1  \\
\hline \label{nl}
\end{tabular}
\end{table}

Sidebands were assumed to lie at a fixed spacing of 173.6 $\mu$Hz and to have the same width as the main peaks.
The fractional height of the sidebands were also fixed according to the ratio of the sidebands to the main peak
in the fourier transform of the window function. In this way, we did not fit any parameters associated solely
with the sidebands.

The number of fitted parameters was further reduced by assuming fixed height ratios for the rotationally split
components. These were fixed according to equations~\ref{VisInc1}~to~\ref{VisInc5}, again assuming an
inclination for $\alpha$ Cen A of $79^\textup{\footnotesize{o}}$.

Finally, it should be noted that because of the relatively short duration of the time series and the poor SNR,
we did not feel that the data justified the use of the more complicated asymmetric \cite{Nigam1998} model.

To fit the model to the power spectrum we used a Powell multi-dimensional hill-climbing minimization algorithm,
which maximized an appropriate log-likelihood function. This function was based on the assumption that the power
spectrum is distributed with negative exponential (i.e., $\chi^2$, with two degrees of freedom) statistics. The
following parameters were varied iteratively until they converged on their best fitting values.
\begin{enumerate}
\item A frequency for each mode. \item A single height for each mode. Heights of rotationally split components
were fixed relative to the strongest outer $\ell$ = $\mid$$m$$\mid$ components (see
equations~\ref{VisInc1}~to~\ref{VisInc5}). \item A single width for all modes. \item A single splitting for the
$\ell$ = 1 and 2 modes \item A flat, background offset for the whole fit.
\end{enumerate}
A total of 2$M$+3 parameters were therefore fitted where $M$ is the number of modes included in the fitting
(i.e, 18 in this case). We should add that in order to recover the power, (i.e., the square of the amplitude),
$A^2$, in each mode from the fitted height, $H$, and the width, $\Delta$, we used the following expression:
\begin{equation}
A^2 = H \left(\frac{\pi}{2} T \Delta + 1 \right),
\label{PowHeightMod1}
\end{equation}
where $T$ is the length of the time series. Details on the derivation of this expression are given in the
appendix.

\subsection{The Auto-Covariance Function}\label{ACF_Sec}

The Autocovariance Function (ACF) is the product of a data series with a shifted versions of itself over
successive time lags, $\tau$. For a time series of discrete measurements, $x_i$, it is defined as:
\begin{equation}
Y_\tau = \frac {\sum_{i=0}^{(N-1)-\tau} \mbox{ } (x_i -
\overline{x}) (x_{i+\tau} - \overline{x})} {N-\tau}.
\end{equation}
where $N$ is the number of points in the data series and $\overline{x}$ is the mean of the sample. In many data
series the mean is often very close to zero, as is the case for p-mode intensity or Doppler velocity residuals,
in which case the ACF reduces to:
\begin{equation}
Y_\tau = \frac {\sum_{i=0}^{(N-1)-\tau} \mbox{ } x_i x_{i+\tau}}
{N-\tau}. \label{ACFeq}
\end{equation}

The more commonly used normalized version of the autocovariance function is termed the autocorrelation function
and for a time series with zero mean is given by:
\begin{equation}
\rho_\tau = \frac {\sum_{i=0}^{(N-1)-\tau} \mbox{ } x_i
x_{i+\tau}} {N-\tau \sum_{i=0}^{N-1} \mbox{ } x_i^2}.
\end{equation}
However, by fitting the autocovariance function we can directly obtain absolute estimates of the powers
associated with the modes, as opposed to relative powers that would be obtained from fitting the autocorrelation
function.

Since the ACF is computed in the time domain, the periodic waveforms of the various components are superimposed.
Therefore, to fit a certain set of modes in a given frequency range one must first apply a band-pass filter to
the time series. As the observable modes in the WIRE data lie in the region 1700 $ \leq \nu \leq $ 2650 $\mu$Hz
we need to at least filter over this range. However, we must also take into account the effect of the window
function. As one would expect, the sidebands in the power spectrum manifest in the ACF as additional waveforms.
Hence, we choose to extend the bandpass range to 1500 $ \leq \nu \leq $ 2850 $\mu$Hz in order to include all the
sideband frequencies.

\begin{figure}
\centering \subfigure[ACF -- 10 hours] {\label{ACF-10h}\includegraphics[width=3.3in]{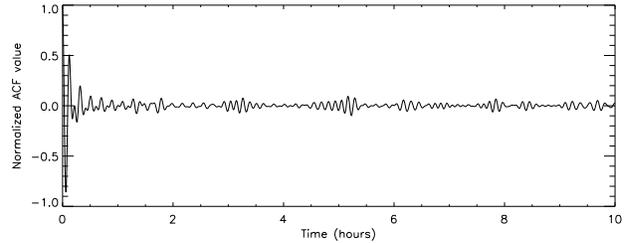}} \subfigure[ACF --
120 hours] {\label{ACF-120h}\includegraphics[width=3.3in]{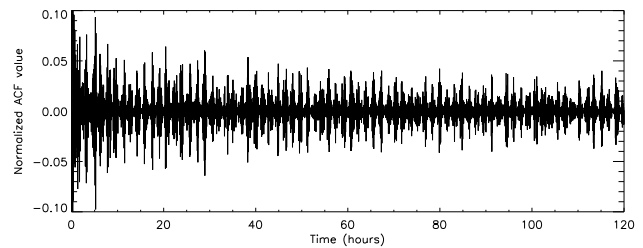}} \caption{Normalized ACF of WIRE data over
(a) the first 10 hours and (b) the first 120 hours. The scale has been reduced in (b) in order to better show
the features.} \label{ACF}
\end{figure}

In Fig.~\ref{ACF} we show the normalized ACF (i.e., the autocorrelation function) of the WIRE time series,
filtered between 1500 $ \leq \nu \leq $ 2850 $\mu$Hz, and plotted over: the first 3600 points (10 hours) in
Fig.~\ref{ACF-10h}; and over the first 43200 points (120 hours) in Fig.~\ref{ACF-120h}. The filtering was
performed using a non-recursive digital filter \citep{Walraven1984}. There are a number of distinctive features
in the ACF the most striking being the very sharp decay in the ACF that occurs over the first few hours. This is
due to the poor SNR in the WIRE data and is explained in more detail below.

After this initial decay the ACF structure is dominated by the p-mode signal. On the smallest timescale, there
is a quasi sinusoidal waveform with a period of about 7.5 minutes. This particular feature can be seen in
Fig.~\ref{ACF-10h} but is too rapid to be seen in Fig.~\ref{ACF-120h}. It is due to the main set of modes that
are seen at around 2000 $\mu$Hz in the $\alpha$-Cen spectrum. On a longer timescale we see quasi-periodic
structure close to $\sim$1.5 hours. This is due to the prominent sidebands which occur at 173.6 $\mu$Hz to
either side of the main peaks. We also see a hint of a $\sim$5-hour quasi-periodic structure due to the
separation between $\ell$ = 0 and $\ell$ = 1 modes. Longer time scale quasi-periodic variations due to smaller
mode spacings, such as those between adjacent modes in the low-$\ell$ pairs, and rotational splitting, are more
difficult to observe and therefore extract. However, there is evidence of an overall exponential decay in the
structure of the ACF due to the finite lifetimes of the modes.

The ACF was modelled using a damped harmonic oscillator equation summed over all visible modes and corresponding
sidebands in the chosen frequency range with an additional background function, $B_\tau$,
\begin{equation}
f_\tau = B_\tau + \sum_{nlmk} A_{nlmk}^2 \cos \left(
\omega_{nlmk}t_\tau \right) \exp(-\gamma t_\tau), \label{Model}
\end{equation}
where,
\begin{displaymath}
\omega_{nlmk} = \sqrt{\left( 2\pi \left( \nu_{nl} + ms + kw
\right)\right)^2 - \gamma^2},
\end{displaymath}
$A^2$ is the mode power, $\omega$ is the natural angular frequency of the mode, $\gamma$ is the damping constant
and $t$ is time. Note that $\gamma$ is related to the width of mode peaks by $\Delta=\gamma/\pi$. $B_\tau$ is
the background component but in the ACF this is not a simple offset but must be treated according to the
expression:
\begin{displaymath}
B_\tau = b \left( \frac {\sin(2 \pi \nu_1 t_\tau)} {2 \pi \nu_1
t_\tau} \right) \cos (2 \pi \nu_2 t_\tau),
\end{displaymath}
where $b$ is the power in the background and is the parameter to be fitted. $\nu_1$ is a value given by the
extent of the filtered frequency range divided by 2 (i.e., 675 $\mu$Hz in this instance) and $\nu_2$ is the
central frequency in the filtered range (i.e., 2175 $\mu$Hz). The sinc term is a direct result of filtering over
a finite frequency range. The wider this range the higher the frequency and the faster the background function
decays. It is this decay that dominates the first few time samples of the WIRE ACF. The extra waveforms due to
the window function are included in the model in the same way as sidebands were treated in the power spectrum.

For the ACF, a gradient-expansion algorithm was used to perform a non-linear least-squares fit to our model,
with
\begin{equation}
\chi^2 = \sum_\tau \frac {[Y_\tau-f(t_\tau,\textbf{a})]^2}
{\sigma_\tau^2} \label{LS}
\end{equation}
the quantity to be minimized, where $Y$ are the data, $f$ is the model, $t$ is  time, $\textbf{a}$ is the vector
of parameters and $\sigma$ is the expected error on each point of the ACF. Note that we are essentially fitting
the same set of parameters as with the power spectrum except the power (square of the amplitude) in the modes is
fitted directly.

We note that two assumptions are made in order to simplify fitting the model to the ACF. Firstly, we take the
error distribution, $\sigma_\tau$ to be constant over the range of the ACF that we fit. A plot of the standard
deviation over all the ACF's of the simulated data shows this to be a reasonable approximation. Secondly, we
ignore the effect of correlation between one point in the ACF and the next. While correlation clearly must be
present, the effect of ignoring it can be shown to only affect attempted error calculations and not the fitted
values themselves \citep{Kuan2000}; hence the reason for using Monte Carlo simulations to fix the errors on our
fits.

\section{Results and Analysis}\label{Results_Sec}

In this section we present the results and analysis of fitting the WIRE data using the two methods described in
Section~\ref{ModelFit_Sec}. We initially show a graphical representation of the fitted model overlaid on the
power spectrum before analyzing the four different types of parameters fitted (frequencies, linewidth,
amplitudes and background) in more detail in separate subsections.

\subsection{Graphical Representation of the Fits}\label{Graph_Sec}

\begin{figure*}
\centerline{\includegraphics[width=7.0in]{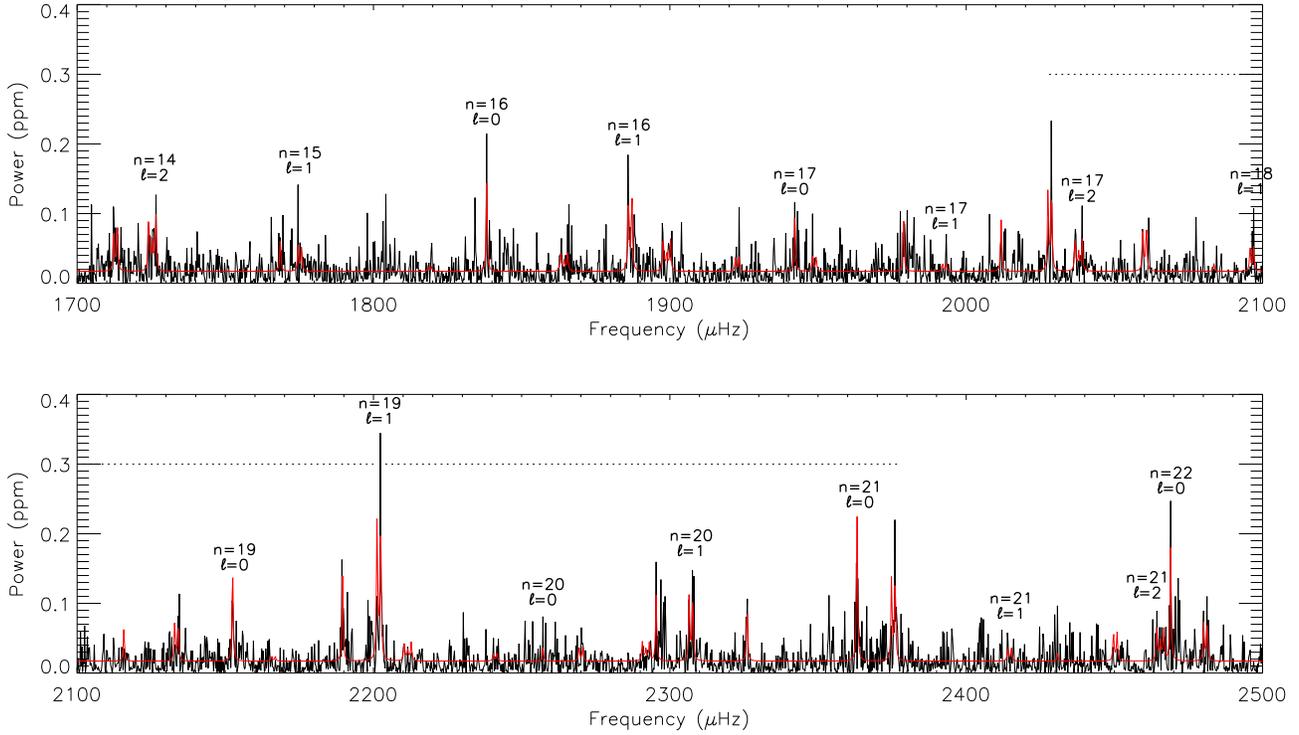}}\caption{WIRE Spectrum for the region 1700 $
\leq \nu \leq $ 2500 $\mu$Hz, with the fitted model overlaid . The fitted modes are labelled and the horizontal
lines extending from the strong peak at $\sim$ 2200 $\mu$Hz indicate the scale of the sideband distance. Three
of the modes have very questionable fits, those at $n$=17, $\ell$=1, $n$=20, $\ell$=0 and $n$=21, $\ell$=1.}
\label{Wire_PowSpec_FitAll}
\end{figure*}

In order to give a clear picture of the fits as determined from the power spectrum fitting technique,
Fig.~\ref{Wire_PowSpec_FitAll} shows the fitted model overlaid on the WIRE power spectrum in the region 1700 $
\leq \nu \leq $ 2500 $\mu$Hz. This plot shows a very good illustration of the fitted splittings. (Note that in
sections~\ref{RotSplit_Sec} and \ref{Width_Sec} we show that fits to the power spectrum have a tendency to
slightly overestimate the splitting and underestimate the width.)

\begin{figure*}
\centerline{\includegraphics[width=7.0in]{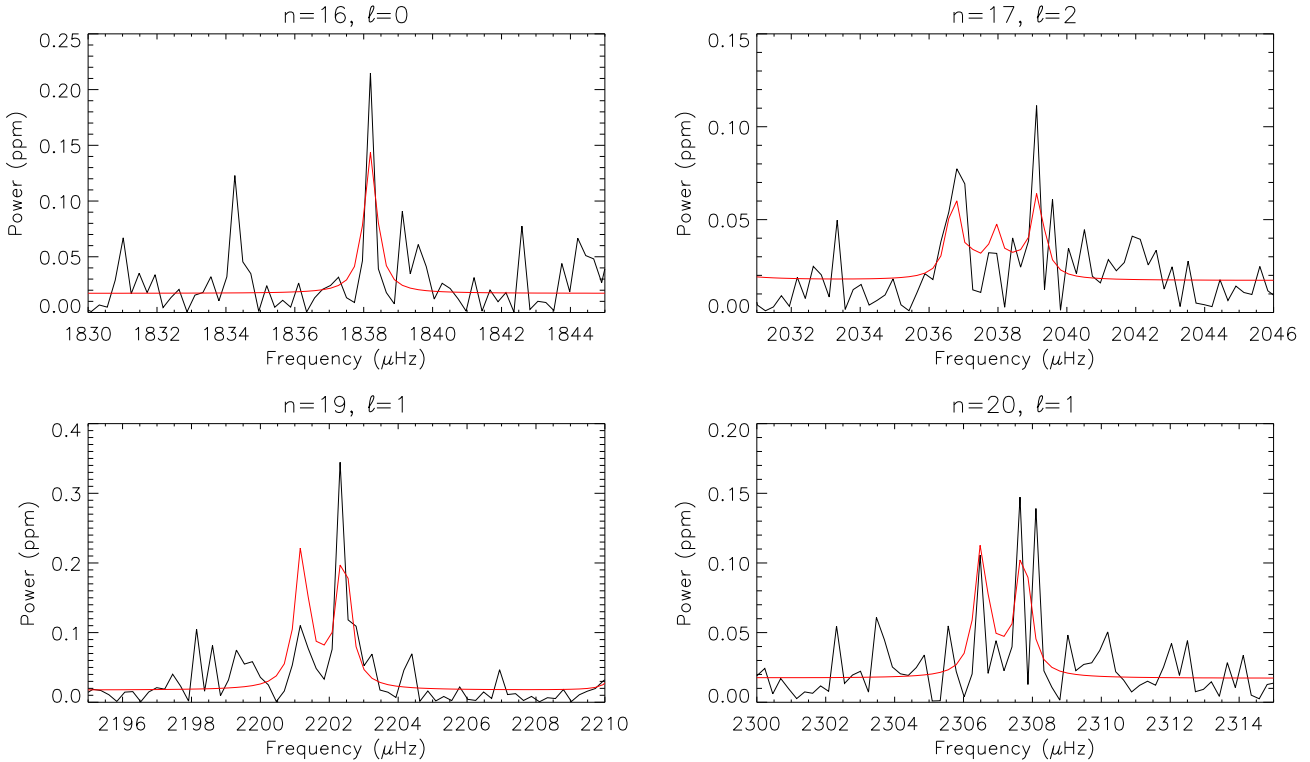}}\caption{15 $\mu$Hz slices of the WIRE spectrum
showing the fits for four different modes.} \label{Wire_PowSpec_FitSep}
\end{figure*}

The fitted peaks that are not labelled in the plot are the first order sidebands. The dotted line extending
outwards from the strong $n$=19, $\ell$=1 peak shows the frequency spacing at which the associated sidebands are
placed. Having sidebands occur at such a large distances from their main peak is somewhat unusual for those more
used to dealing with diurnal sidebands.

On closer inspection of the plot, there are some fitted modes that have very small powers (namely the $n$=17,
$\ell$=1, $n$=20, $\ell$=0 and the $n$=21, $\ell$=1 modes) and as such, the reliably of these fits is somewhat
questionable. The reason for their presence is that the fitting routine attempts to fit peaks for all the modes
included in the model. Hence, if a mode is expected at a certain frequency, and there is no large and obvious
peak in the nearby vicinity, the fitting routine may latch on to a smaller peak which is only associated with
the background noise. However, with these fitted modes having such small powers, even if they are not true
signatures of mode excitation, their effect on the overall width and rotational splitting estimates is
negligible.

In Fig.~\ref{Wire_PowSpec_FitSep}, four 15 $\mu$Hz slices of the spectrum are shown, each located on fairly
strong modes. For the $\ell$=1 and $\ell$=2 modes, it is possible to identify a rotational splitting pattern in
the spectrum, and the model is seen to fit reasonably well. While we would also like to be able to show similar
graphical illustrations of the splitting and widths as determined from the ACF fitting, the superimposed nature
of the waveforms makes this difficult.

\subsection{Mode Frequencies}\label{Freq_Sec}

Both methods produced well-constrained frequency estimates for seventeen out of eighteen of the modes we tried
to fit. Sixteen of these were determined via both methods while the other two modes were fitted by one method
only. The fitted frequencies are shown in Tables~\ref{freqPS} and \ref{freqACF} respectively where the quoted
uncertainties were determined from the standard deviation of fits made to the 50 simulated spectra. The errors
are quite large for some of the estimates due to the high background and resulting poor SNR of the WIRE data. A
comparison between the two sets of results shows the differences to be very small. The root-mean-square (rms)
difference is 0.3 $\mu$Hz, compared with an average error of 1.4 $\mu$Hz.

We can also compare our frequencies with those determined by \cite{Bedding2004} and \cite{Bouchy2002}. Looking
at individual modes we see differences of the order of 1 $\mu$Hz which is within the errors given. However, when
taking a weighted average of the differences we find the estimates from the WIRE data are about 0.8 $\pm$ 0.3
$\mu$Hz lower than the frequencies determined from \cite{Bedding2004} and \cite{Bouchy2002}. One contribution to
this difference comes from the shift in the mode frequencies due to the relative motion of the Earth around its
orbit. By considering the times of year during which the observations were carried out, and the ecliptic
coordinates of $\alpha$ Cen A, it can be shown that this effect should lead to the WIRE data having frequencies
$\sim$ 0.2 $\mu$Hz lower than the other two data sets. Therefore, the corrected difference is 0.6 $\pm$ 0.3
$\mu$Hz, which is only significant at 2 sigma.

Since the WIRE data set is taken 19 months prior to the others one might conjecture that the difference is due
to an activity cycle on $\alpha$ Cen A. Were a 0.6 $\mu$Hz shift over a period of just 19 months to be real, it
would suggest a large and rapid (or possibly just very large) activity cycle. By comparison, the minimum to
maximum change in mode frequencies due to the solar cycle, is $\sim$0.4 $\mu$Hz and this occurs over a 5.5-year
period. Of course, a zero or very small shift is not ruled out and so it is clear that further and better
quality sets of observations would be needed in order to make any solid conclusions.

\begin{figure}
\centering \subfigure[Power Spectrum] {\label{FreqDiff_PS}\includegraphics[width=1.6in]{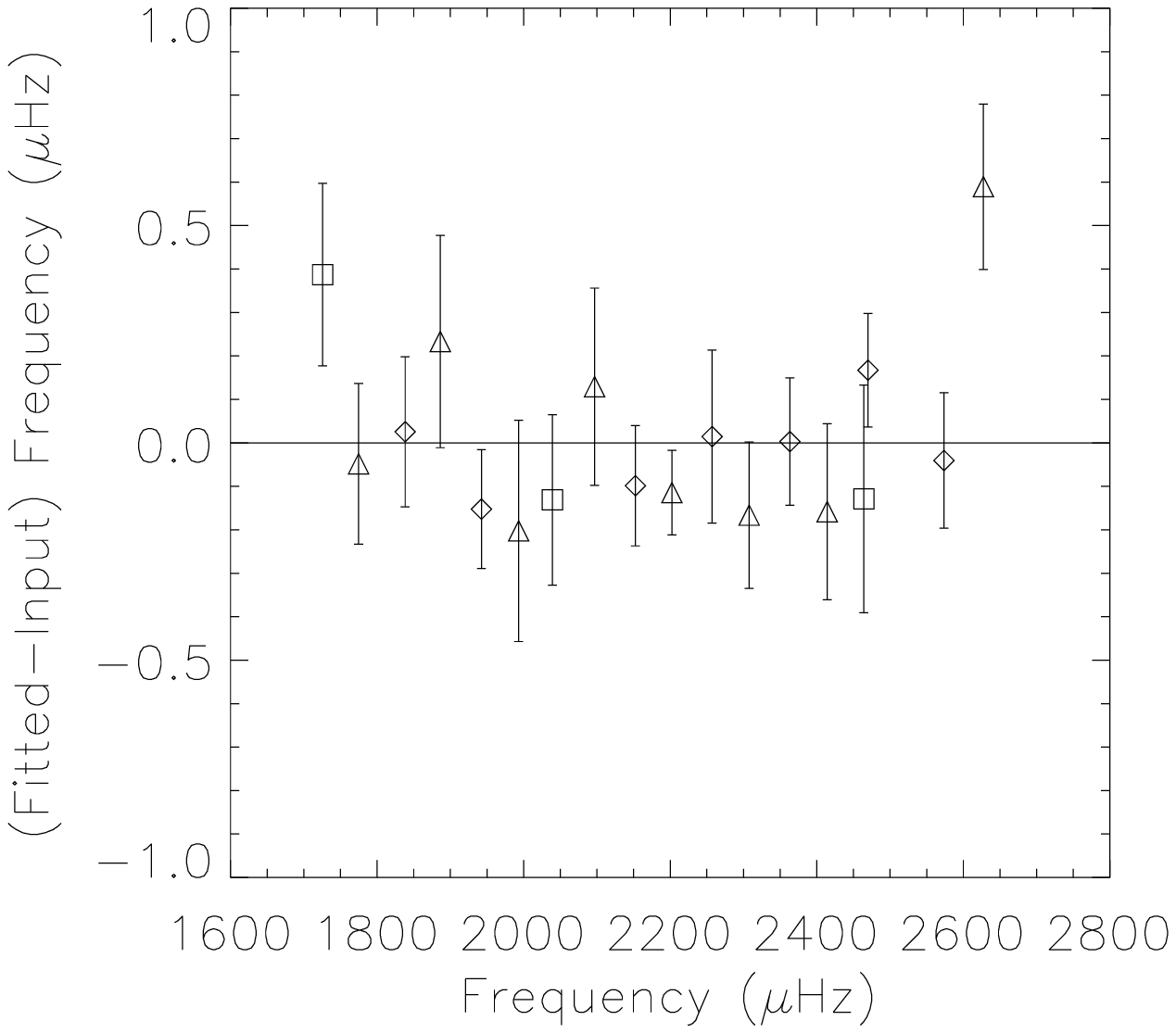}}
\subfigure[ACF] {\label{FreqDiff_ACF}\includegraphics[width=1.6in]{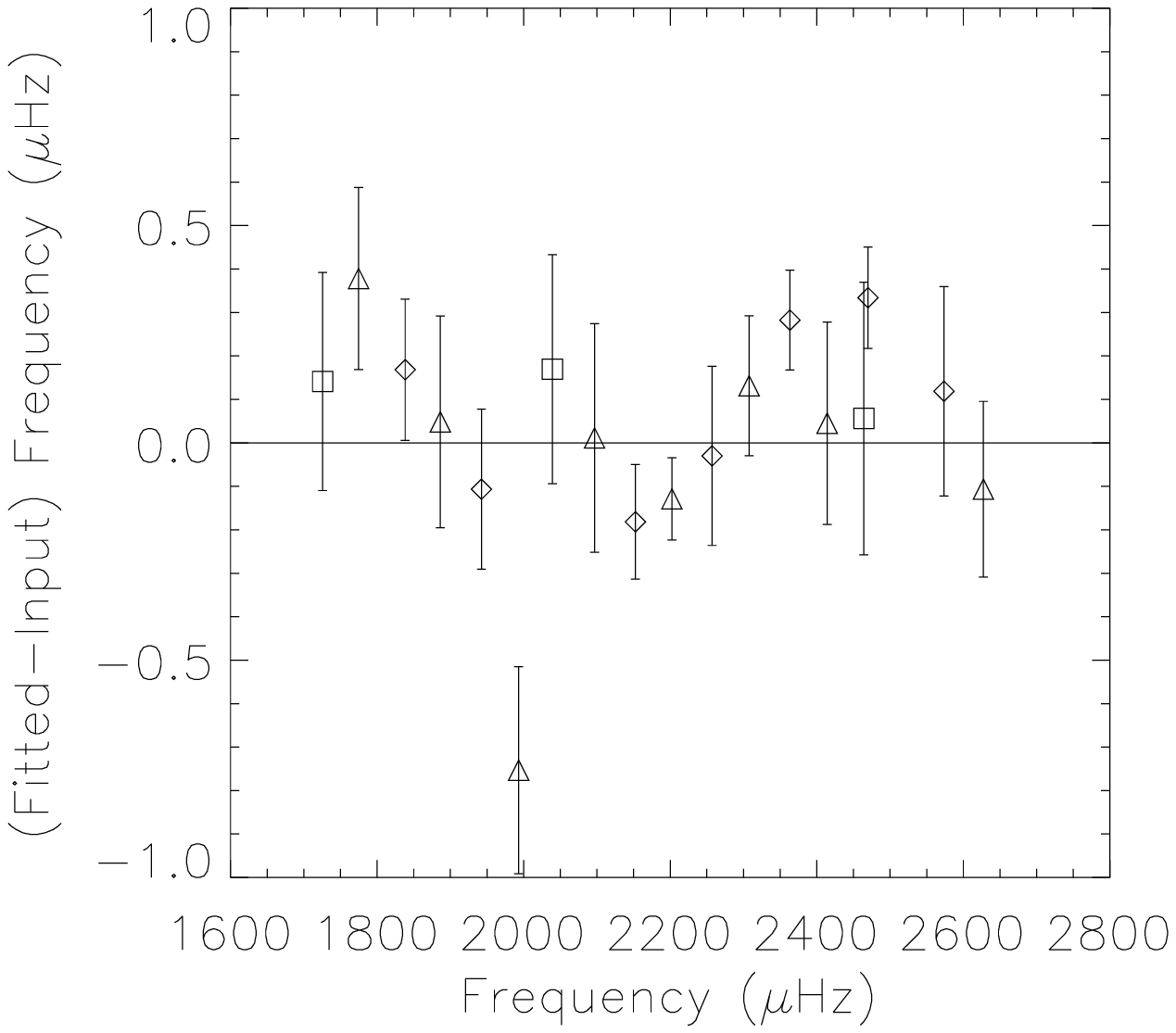}} \caption{Results of fitting
artificial WIRE-like data. Average difference between fitted and input frequencies for $\ell$=0 (diamonds),
$\ell$=1 (triangles) and $\ell$=2 (squares) modes. Error bars indicate the error on the mean as given by the
standard deviation in the fitted frequencies divided by the square root of the number of fits. \vspace{2mm}}
\label{FreqDiff}
\end{figure}

As well as using the simulated data to estimate the uncertainties on our fits we also used them to test accuracy
and reliability. The plots in Fig.~\ref{FreqDiff} show the mean difference between the fitted and input
frequencies of the simulations. The associated error bars are the error on the mean, given by the standard
deviation divided by the square root of the number of fits. For both methods the estimates are generally
accurate to within errors and there does not seem to be any particular bias towards an under or overestimation
of the input frequency.

\begin{table}
\caption{Frequencies extracted by power spectrum fitting method ($\mu$Hz). Brackets indicate modes with
questionable fits, see Section~\ref{Graph_Sec}} \centering
\begin{tabular}{cccc}
\hline $n$&$\ell$=0&$\ell$=1&$\ell$=2\\ \hline
14&                  &                  & 1725.4 $\pm$ 1.5 \\
15&                  & 1775.2 $\pm$ 1.3 &                  \\
16& 1838.2 $\pm$ 1.2 &[1886.6 $\pm$ 1.7]&                  \\
17& 1942.4 $\pm$ 1.0 & 1992.7 $\pm$ 1.8 & 2037.9 $\pm$ 1.4 \\
18&                  & 2096.4 $\pm$ 1.6 &                  \\
19& 2152.5 $\pm$ 1.0 & 2201.8 $\pm$ 0.7 &                  \\
20&[2257.2 $\pm$ 1.4]& 2307.1 $\pm$ 1.2 &                  \\
21& 2363.2 $\pm$ 1.0 &[2414.6 $\pm$ 1.4]& 2465.5 $\pm$ 1.8 \\
22& 2469.0 $\pm$ 0.9 &                  &                  \\
23&                  & 2623.9 $\pm$ 1.3 &                  \\
\hline \label{freqPS}
\end{tabular}
\end{table}

\begin{table}
\caption{Frequencies extracted by ACF fitting method ($\mu$Hz).} \centering
\begin{tabular}{cccc}
\hline $n$&$\ell$=0&$\ell$=1&$\ell$=2\\ \hline
14&                  &                  & 1725.3 $\pm$ 1.8 \\
15&                  & 1774.9 $\pm$ 1.5 &                  \\
16& 1838.5 $\pm$ 1.2 & 1886.6 $\pm$ 1.7 &                  \\
17& 1942.4 $\pm$ 1.3 &[1993.1 $\pm$ 1.8]& 2038.8 $\pm$ 1.9 \\
18&                  &[2096.9 $\pm$ 1.9]&                  \\
19& 2152.5 $\pm$ 0.9 & 2202.1 $\pm$ 0.7 &                  \\
20&[2257.1 $\pm$ 1.5]& 2307.7 $\pm$ 1.1 &                  \\
21& 2363.4 $\pm$ 0.8 &                  & 2465.6 $\pm$ 2.2 \\
22& 2469.2 $\pm$ 0.8 &                  &                  \\
23& 2578.3 $\pm$ 1.7 & 2624.3 $\pm$ 1.4 &                  \\
\hline \label{freqACF}
\end{tabular}
\end{table}

The errors on the real data estimates given in Tables~\ref{freqPS} and \ref{freqACF} and on the artificial data
estimates in Fig.~\ref{FreqDiff} show how the uncertainties vary quite substantially from one mode to the next.
This is to be expected since the uncertainties depend strongly on the SNR, when the SNR is poor. The equation
(\citeauthor{Libbrecht1992} \citeyear{Libbrecht1992}; \citeauthor{Toutain1994} \citeyear{Toutain1994}):
\begin{equation}
\sigma^2_{\nu_{nl}} = \frac {\Delta} {4 \pi T} \sqrt{\beta+1}
(\sqrt{\beta+1}+\sqrt{\beta})^3,
\end{equation}
gives an estimate of the error on the fitted frequency, $\sigma_{\nu_{nl}}$ as a function of the linewidth,
$\Delta$, the length of the time series, $T$ and the noise-to-signal ratio, $\beta$ (i.e the inverse of the
SNR). When the SNR is high, $\beta$ is small, and the level of uncertainty will be dominated by $\Delta$ and $T$
and as such will remain fairly constant from mode to mode. However, for a poor SNR, the $\beta$ term is
important and so small changes in this, due to the difference in amplitudes of the modes, will have a large
effect on the uncertainty.

\begin{figure}
\centering \subfigure[Power Spectrum] {\label{FreqErr-Power_PS}\includegraphics[width=1.6in]{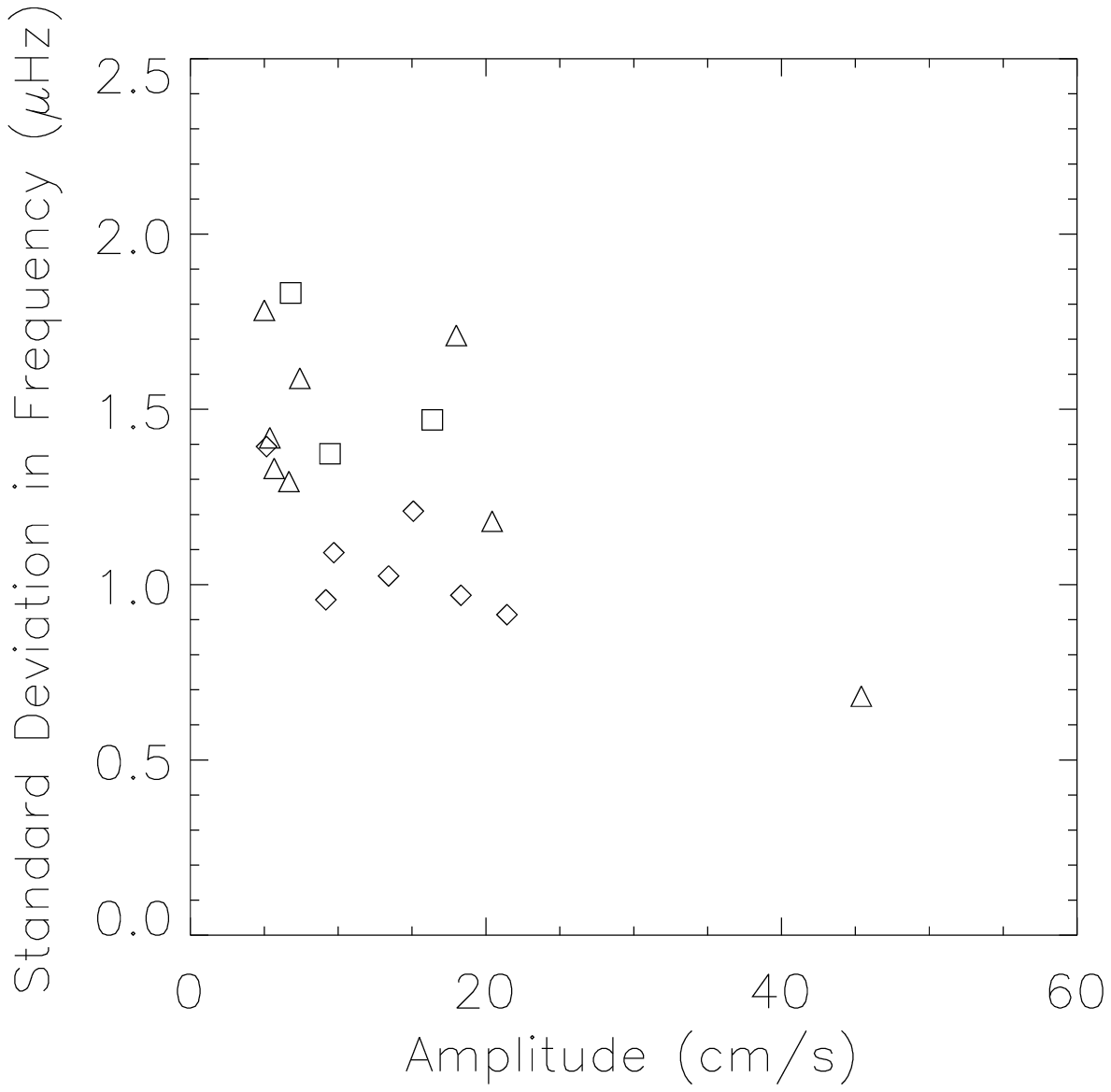}}
\subfigure[ACF] {\label{FreqErr-Power_ACF}\includegraphics[width=1.6in]{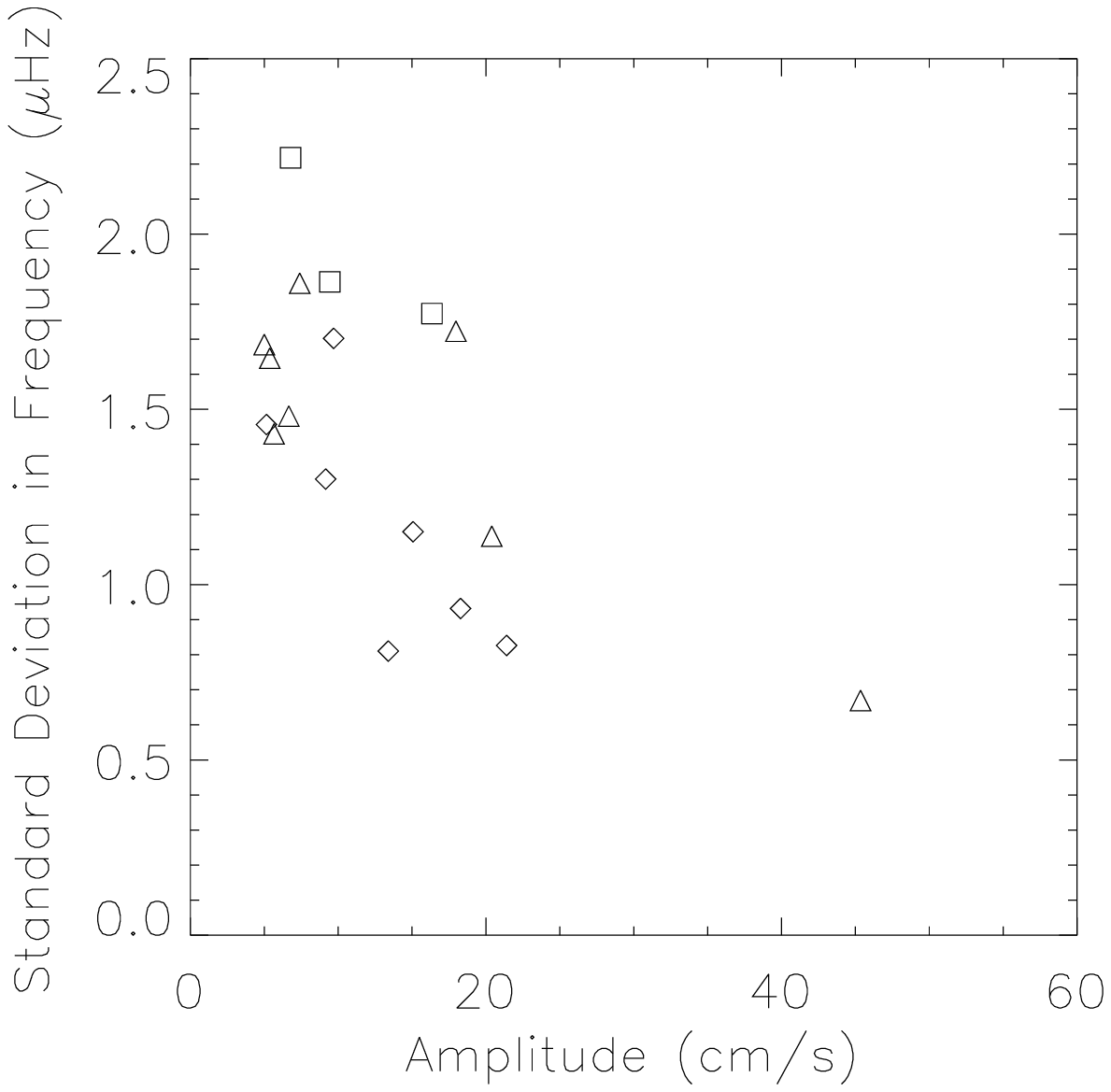}} \caption{Results of
fitting artificial WIRE-like data. Standard deviation of fitted frequencies as a function of the input amplitude
of the modes. Symbols have the same meanings as in Fig.~\ref{FreqDiff}.\vspace{2mm}} \label{FreqErr-Power}
\end{figure}

The relationship between the SNR and the errors on the frequencies can be seen more clearly by plotting the
standard deviation against the input amplitude of the modes, as in Fig.~\ref{FreqErr-Power} (this is equivalent
to plotting against the signal-to-noise in amplitude since the background is kept constant with frequency). The
large SNR in the WIRE data also means that the typical uncertainties on our frequency estimates are somewhat
larger than those given by \cite{Bedding2004}, which were $\sim$0.3 $\mu$Hz.

\subsection{Rotational Splitting}\label{RotSplit_Sec}

There have to date been no successful attempts to measure the rotational splitting of the p-modes in $\alpha$
Cen A and thereby make a first step to constraining the internal rotation. However, by applying the ACF fitting
technique to the WIRE data and assuming a constant splitting across the frequency range we have been able to
determine a reasonably well constrained estimate of 0.54 $\pm$ 0.22 $\mu$Hz. The simulated data were again used
in order to check the accuracy of this estimate. By fitting the ACF of the 50 artificial data sets an average
splitting of 0.44 $\pm$ 0.03 $\mu$Hz was returned. This was within 2 standard errors of the input value of 0.5
$\mu$Hz, indicating that the fits are relatively unbiased. It should be noted that of the 50 estimates returned
via the ACF method, 10 gave a splitting very close to zero. These `null' splitting results are a common problem
when fitting poorly defined modes (either due to poor resolution, low SNR or a large width-to-splitting ratio),
and we have taken the usual step of disregarding these values when determining the mean and standard deviation
of the fits to the simulated data \citep{Chaplin2001}.

A value of the splitting was also returned when fitting the power spectrum. This method gave an estimate of 0.64
$\pm$ 0.25 $\mu$Hz, which is within errors of the estimate given by fitting the ACF. However, fits to the
simulated data showed a tendency for the estimates of the splitting to overestimate the true input value. An
average of 0.68 $\pm$ 0.03 was returned which gives a result which is $\sim$6 standard errors too large. This is
also a common problem in the solar case but is generally due to a large width-to-splitting ratio at high
frequencies rather than a poor SNR. Because of the bias in the power spectrum estimate we would suggest that the
value of the rotational splitting returned via fitting the ACF is more robust.

A rotational splitting of 0.54 $\pm$ 0.22 $\mu$Hz gives a rotational period of 21$^{+17}_{-5}$ days. This agrees
with a measurement of the $\alpha$ Cen A surface rotation rate determined by \cite{Jay1996} of 23$^{+5}_{-2}$
days using chromospheric and transition region emission line features as markers of rotation. Also, a surface
rotation rate of 2700 $\pm$ 700 ms$^{-1}$ from \cite{Saar1997} coupled with theoretical predictions of the alpha
Cen A radius of 1.23 R$_\odot$ \citep{Morel1995}, give a prediction of 23 $\pm$ 6 days. A similar correspondence
of the low-$\ell$ splittings and the surface rate of rotation is seen for the Sun.

\subsection{Linewidths and Lifetime}\label{Width_Sec}

Since our model also assumed a fixed value for the linewidth we were able to obtain a well constrained estimate
of this parameter. However we lose any knowledge of how the parameter varies with frequency. Our estimate for
the average linewidth of the modes from fitting the ACF of the WIRE time series is 0.92 $\pm$ 0.30 $\mu$Hz,
which equates to a lifetime of 3.9 $\pm$ 1.4 days. Fits to the ACF of the simulated data gave an average width
of 1.10 $\pm$ 0.05 $\mu$Hz, which is again within 2 standard errors of the input value of 1.0 $\mu$Hz. Our fits
to the artificial data also showed a clear anti-correlation between the fitted linewidths and the fitted
rotational splittings. This meant that the fits that returned `null' splittings also returned artificially high
linewidths, as one might expect. We therefore removed these fits from our calculation of the mean and standard
deviation of the artificial results, in the same manner as for the rotational splitting.

Our estimates of the linewidth from fitting the power spectrum gave us a much smaller value of 0.46 $\pm$ 0.38
$\mu$Hz, which is a lifetime of 8.1 $\pm$ 6.8 days. However, fits to the simulated data showed estimates of the
linewidth to be significantly underestimated. An average fitted value across the 50 data sets of 0.62 $\pm$ 0.06
$\mu$Hz was returned indicating an underestimate of over 6 standard errors. The same anti-correlation between
linewidth and rotational splitting values that was seen when fitting the ACF was also seen here with
overestimates of the splitting being combined with underestimates in the linewidth. This leads us to conclude
that the ACF fitting technique also gives a more reliable estimate of the linewidth and hence lifetime of the
$\alpha$ Cen A p-modes.

In relation to our assumption that fitting a constant width returns the average linewidth across the fitted
modes we note that the ACF fitting has also been tested on artificial data simulating solar oscillations, for
which the linewidth varies with frequency. In this case we found that fitting a constant width did indeed return
the input average across the modes.

The \cite{Bouchy2002} and \cite{Bedding2004} studies were both made from observations lasting just a few days (5
and 12 respectively) and a direct measurement of the mode lifetime was difficult to achieve. However,
\cite{Bedding2004} were able to develop a method of estimating the lifetime from the scatter in the frequency
measurements. Their initial estimate put the average mode lifetime at $1.4^{+0.5}_{-0.4}$ days at 2.1 mHz,
however, that has recently been reevaluated as $2.3^{+1.0}_{-0.6}$ days \citep{Kjeldsen2005}. This estimate is
smaller than our value determined from fitting the ACF of the WIRE data, although the error bars overlap.

\subsection{Mode Amplitudes} \label{Amps_Sec}

\begin{table}
\caption{Amplitudes extracted by power spectrum fitting method (cms$^{-1}$). Brackets indicate modes with
questionable fits.} \vspace{0.0mm} \centering
\begin{tabular}{@{\extracolsep{4.0mm}}cccc}
\hline \hspace{3mm}\small{$n$}&\small{$\ell$=0}&\small{$\ell$=1}&\small{$\ell$=2}\\ \hline \vspace{-3.0mm} \\
\hspace{3mm}\small{14}& & &\scriptsize{$41^{+34}_{-14}$}\\
\hspace{3mm}\small{15}& &\scriptsize{$13^{+18}_{-\hspace{3.2pt}5}$}& \\
\hspace{3mm}\small{16}&\scriptsize{$23^{+21}_{-\hspace{3.2pt}8}$}&\scriptsize{$41^{+34}_{-12}$}& \\
\hspace{3mm}\small{17}&\scriptsize{$14^{+12}_{-\hspace{3.2pt}5}$}&[\hspace{4pt}\scriptsize{$4^{+13}_{-\hspace{3.2pt}4}$}\small{ ]}&\scriptsize{$23^{+20}_{-\hspace{3.2pt}9}$}\\
\hspace{3mm}\small{18}& &\scriptsize{$13^{+19}_{-\hspace{3.2pt}6}$}& \\
\hspace{3mm}\small{19}&\scriptsize{$24^{+26}_{-\hspace{3.2pt}8}$}&\scriptsize{$76^{+54}_{-30}$}& \\
\hspace{3mm}\small{20}&[\hspace{4pt}\scriptsize{$3^{+15}_{-\hspace{3.2pt}4}$}\small{ ]}&\scriptsize{$35^{+38}_{-14}$}& \\
\hspace{3mm}\small{21}&\scriptsize{$38^{+14}_{-\hspace{3.2pt}7}$}&[\hspace{4pt}\scriptsize{$7^{+17}_{-\hspace{3.2pt}4}$}\small{ ]}&\scriptsize{$24^{+15}_{-\hspace{3.2pt}7}$}\\
\hspace{3mm}\small{22}&\scriptsize{$29^{+34}_{-13}$}& & \\
\hspace{3mm}\small{23}& &\scriptsize{$26^{+19}_{-\hspace{3.2pt}6}$}& \\
\hline \label{ampPS}
\end{tabular}
\end{table}

\begin{table}
\caption{Amplitudes extracted by ACV fitting method (cms$^{-1}$).} \vspace{0.0mm} \centering
\begin{tabular}{@{\extracolsep{1.5mm}}ccccc}
\hline \hspace{3mm}\small{$n$}&\small{$\ell$=0}&\small{$\ell$=1}&\small{$\ell$=2}\\ \hline \vspace{-3.0mm} \\
\hspace{3mm}\small{14}&\small{                      }&\small{                      }&\small{25 $\pm$12            }\\
\hspace{3mm}\small{15}&\small{                      }&\small{12 $\pm$ \hspace{1pt}7}&\small{                      }\\
\hspace{3mm}\small{16}&\small{24 $\pm$ \hspace{1pt}8}&\small{25 $\pm$16            }&\small{                      }\\
\hspace{3mm}\small{17}&\small{17 $\pm$ \hspace{1pt}9}&\small{[ 4 $\pm$ \hspace{1pt}7 ]}&\small{19 $\pm$ \hspace{1pt}9}\\
\hspace{3mm}\small{18}&\small{                      }&\small{[ 6 $\pm$ \hspace{1pt}8 ]}&\small{                      }\\
\hspace{3mm}\small{19}&\small{24 $\pm$10            }&\small{52 $\pm$20            }&\small{                      }\\
\hspace{3mm}\small{20}&\small{[ 4 $\pm$ \hspace{1pt}7 ]}&\small{22 $\pm$13            }&\small{                      }\\
\hspace{3mm}\small{21}&\small{29 $\pm$ \hspace{1pt}8}&\small{                      }&\small{17 $\pm$ \hspace{1pt}8}\\
\hspace{3mm}\small{22}&\small{23 $\pm$14            }&\small{                      }&\small{                      }\\
\hspace{3mm}\small{23}&\small{13 $\pm$ \hspace{1pt}8}&\small{19 $\pm$ \hspace{1pt}6}&\small{                      }\\
\hline \label{ampACF}
\end{tabular}
\end{table}

We have chosen to present the strength of the modes in terms of their amplitudes rather than power. We have also
converted the units to velocity (cms$^{-1}$) even though the WIRE observations were intensity measurements. In
both cases this was done so as to more easily make comparisons with the \cite{Bedding2004} study. To convert
between intensity given in parts per million (ppm) and velocity we use the expression given in
\cite{Kjeldsen1995}:
\begin{equation}
v_{\textup{\scriptsize{osc}}} = \left( \frac {\delta L} {L} \right)_{\scriptsize{\lambda}} \left( \frac {\delta
L} {550\textup{nm}} \right) \left( \frac {T_{\textup{\scriptsize{eff}}}} {5777\textup{K}} \right)^2 \left( \frac
{1} {20.1\textup{ppm}} \right)
\end{equation}
where $v_{\textup{\scriptsize{osc}}}$ denotes the amplitude of velocity oscillations in ms$^{-1}$, $(\delta
L/L)_{\scriptsize{\lambda}}$ the change in intensity at the effective wavelength $\lambda$ ($\sim$ 450nm for
WIRE), ${T_{\textup{\scriptsize{eff}}}}$ the effective temperature of the star ($\sim$ 5770 for $\alpha$ Cen A)
and 5777K is taken as the effective temperature of the Sun.

Of the four types of parameter examined in this paper the amplitudes are the least well constrained. This is
because we have attempted to fit a separate power to each mode, rather than an average value across several
modes as was done with the linewidth and splitting. This results in large uncertainties on the amplitude
estimates as shown in Tables~\ref{ampPS}~and~\ref{ampACF} for the power spectrum and ACF fitting respectively.
Indeed the errors on some of the weaker modes are larger than the actual estimated powers. Hence, we cannot put
much credence in these values as they will be dominated by background noise.

\begin{figure}
\centering \subfigure[Power Spectrum] {\label{PowerDiff_PS}\includegraphics[width=1.6in]{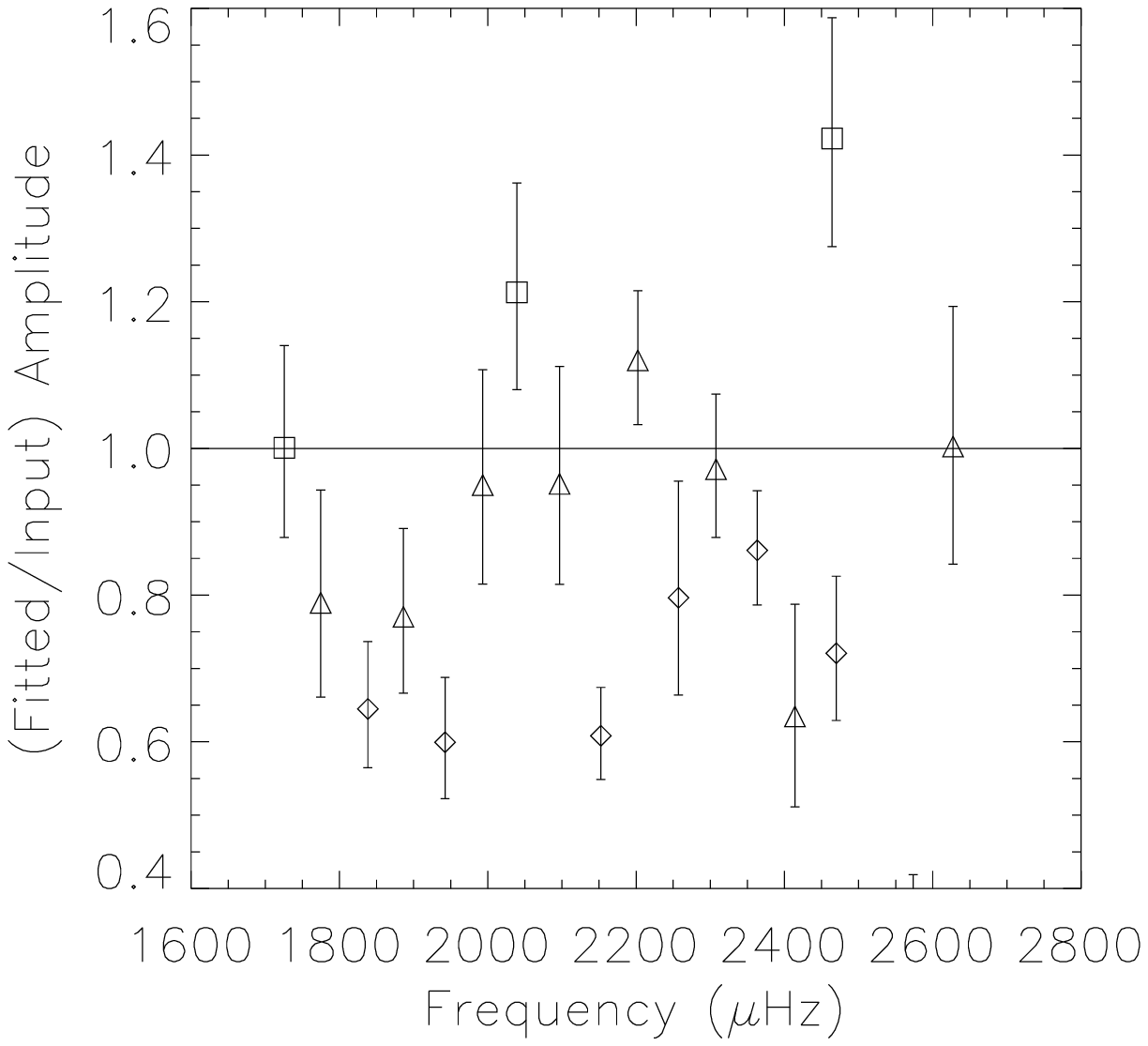}}
\subfigure[ACF] {\label{PowerDiff_ACF}\includegraphics[width=1.6in]{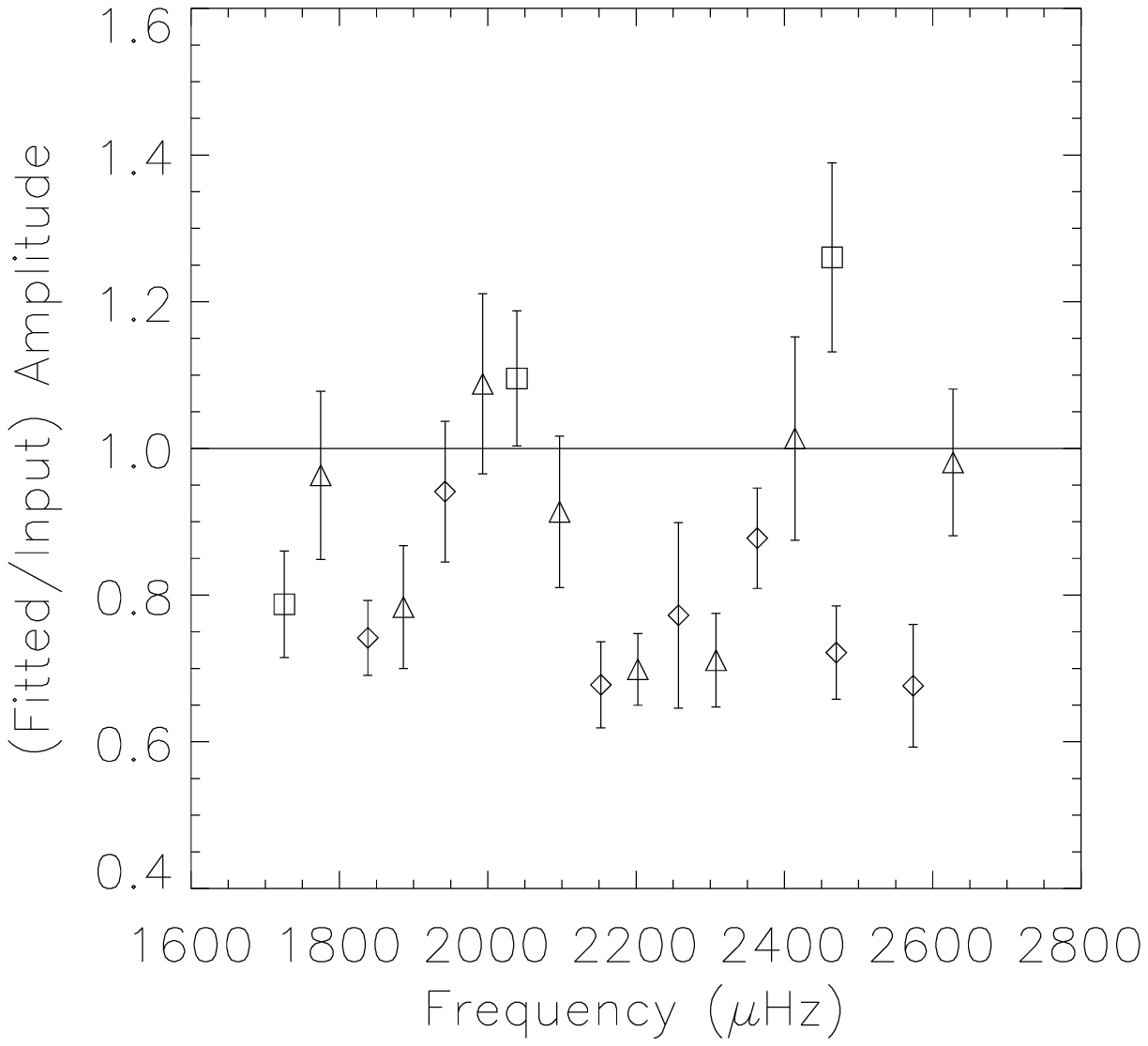}} \caption{Results of
fitting artificial WIRE-like data. Average ratios of fitted to input amplitudes. Symbols have the same meaning
as in Fig.~\ref{FreqDiff}.} \label{PowerDiff}
\end{figure}

\begin{figure}
\centering \subfigure[Power Spectrum]
{\label{PowerDiff-Power_PS}\includegraphics[width=1.6in]{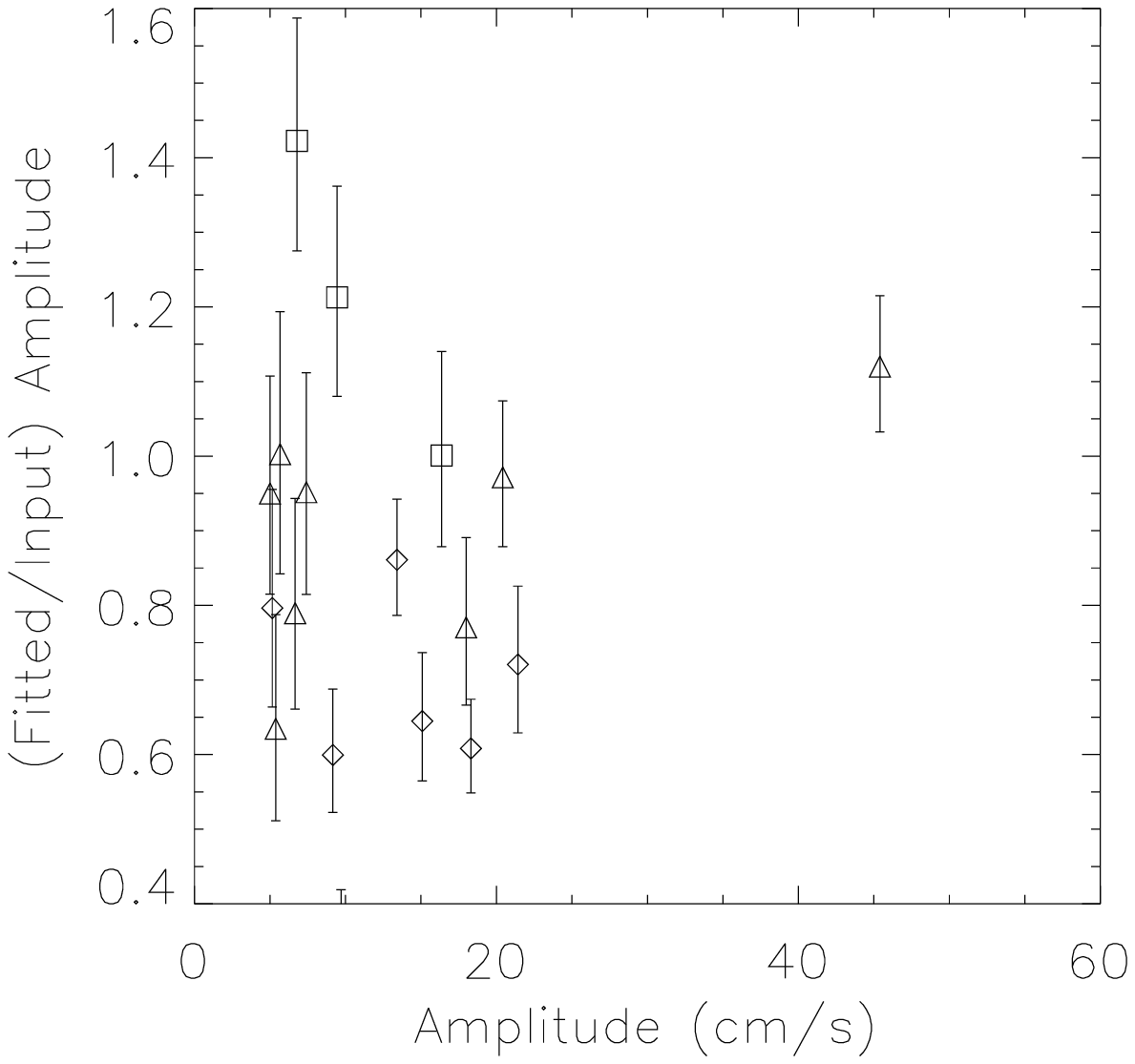}} \subfigure[ACF]
{\label{PowerDiff-Power_ACF}\includegraphics[width=1.6in]{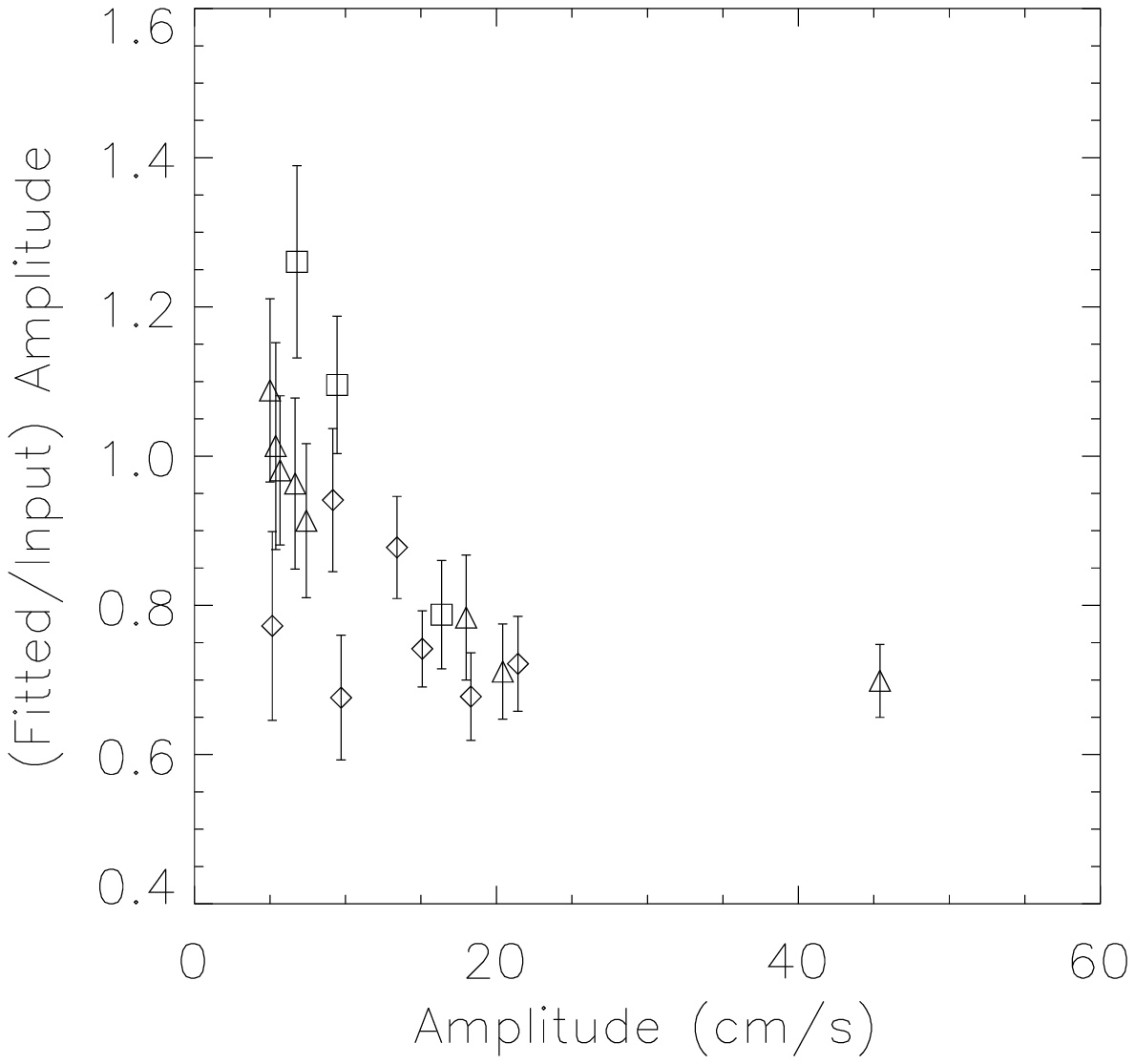}} \caption{Results of fitting
artificial WIRE-like data. Average ratios of fitted to input amplitudes, plotted as a function of input
amplitude. Symbols have the same meaning as in Fig.~\ref{FreqDiff}.} \label{PowerDiff-Power}
\end{figure}

We have again used simulated data to test the accuracy of the fitted powers. In a similar fashion to that used
for the fitted frequencies, we have plotted in Fig.~\ref{PowerDiff} the difference between the power estimates
averaged over the 50 artificial data sets and the input values. The plots show that both methods return a number
of estimates that significantly underestimate the true input amplitude. In Fig.~\ref{PowerDiff-Power} we plot
the same data but as a function of the mode amplitude. This shows that when fitting the ACF, the amplitude of
the stronger modes seem to be underestimated by around 30 percent, whereas the weaker modes suffer less bias.
This pattern is less obvious when fitting the power spectrum with little evidence of the bias being dependent on
the strength of the modes. Fig.~\ref{PowerDiff}, also shows a tendency for the power of the lower $\ell$-valued
modes to underestimate the input values more so than for higher $\ell$-valued modes. This is particulary
noticeable for the power spectrum fits.

\section{Summary}\label{Summary_Sec}

The 50-day time series of photometry observations taken in 1999 by the WIRE spacecraft has been reanalysed using
power spectrum and autocovariance fitting methods. With the help of a-priori information, regarding the location
in frequency of modes from other recent $\alpha$ Cen A studies, we have managed to fit 18 different modes in the
power spectrum and autocovariance function (ACF), 16 of which were fitted in both. The values of the fitted
frequencies are slightly lower than those determined by \citeauthor{Bedding2004} \citeyear{Bedding2004} and
\citeauthor{Bouchy2002} \citeyear{Bouchy2002}, although without better data we cannot say with any confidence
whether this is indicative of an activity cycle for $\alpha$ Cen A.

In addition to the frequencies we have also been able to estimate an average rotational splitting across the
fitted modes of 0.54 $\pm$ 0.22 $\mu$Hz using the ACF. An average lifetime was also estimated by fitting the ACF
and was found to be 3.9 $\pm$ 1.4 days. Although the actual fitted value is larger than the lifetime estimated
by \cite{Kjeldsen2005}, the error bars do overlap. Estimates of the amplitude were also obtained, however they
were rather poorly constrained, especially for the weaker modes.

Simulated time series made to mimic the WIRE data were created in order to test the accuracy and precision of
the fitting methods using a Monte Carlo approach. We found that for the most part, the fitted parameter
estimates averaged over a number of realizations agreed with the input values used to create the data. However,
we did find that fits to the power spectrum tended to underestimate the linewidths and overestimate the
splittings. The bias on both of these parameter estimates were reduced when fitting the ACF.

There still may be opportunities to refine this work further. For example, in this analysis, a fairly basic
approach to dealing with the window function was employed, simply allowing for the subsequent sidebands in the
models. A more sophisticated approach for fitting the power spectrum would be to convolve the spectral window
with the model and fit that to the data. Also, because of the Weiner-Khinchine relation, that states that the
ACF is actually the Fourier transform of the power spectrum, this technique can probably be applied to the ACF
fitting approach as well. Additionally, there is now a new set of WIRE $\alpha$-Cen observations that was taken
in January 2004 and lasted for around 30 days. If modes can be identified and fitted from this data as well, it
will give an excellent comparison with the 1999 time series and should allow for a better investigation into a
possible activity cycle.

\section*{Acknowledgments}

We would like to thank all those who are, or have been, associated with the launching and operation of WIRE. We
also thank the referee, T. Bedding, for his careful review of the paper. STF acknowledges the support of the
School of Physics and Astronomy at the University of Birmingham.

\appendix

\section{Variation of Maximum Power Spectral Density}

Consider a time series of length $T$, of a stochastically excited p-mode with mean power $P$ and a lifetime
$\tau$. The resulting power spectrum peak can then be modelled according to a Lorentzian function with a
linewidth, $\Delta=1/(\pi\tau)$ . Therefore, assuming the modal peak is well resolved in frequency, the maximum
power spectral density per bin, $H$, is given by:
\begin{equation}
H = \frac{2P}{\pi T \Delta} \label{HeightPowW}.
\end{equation}
Notice that $\Delta$ is multiplied by $T$ in order to give the width in terms of the number of bins rather than
Hz. This expression can be rewritten in terms of the lifetime $\tau$.
\begin{equation}
H = 2P \left(\frac{\tau}{T}\right). \label{HeightPowTau}
\end{equation}

\begin{figure}
\centerline{\includegraphics[width=2.8in]{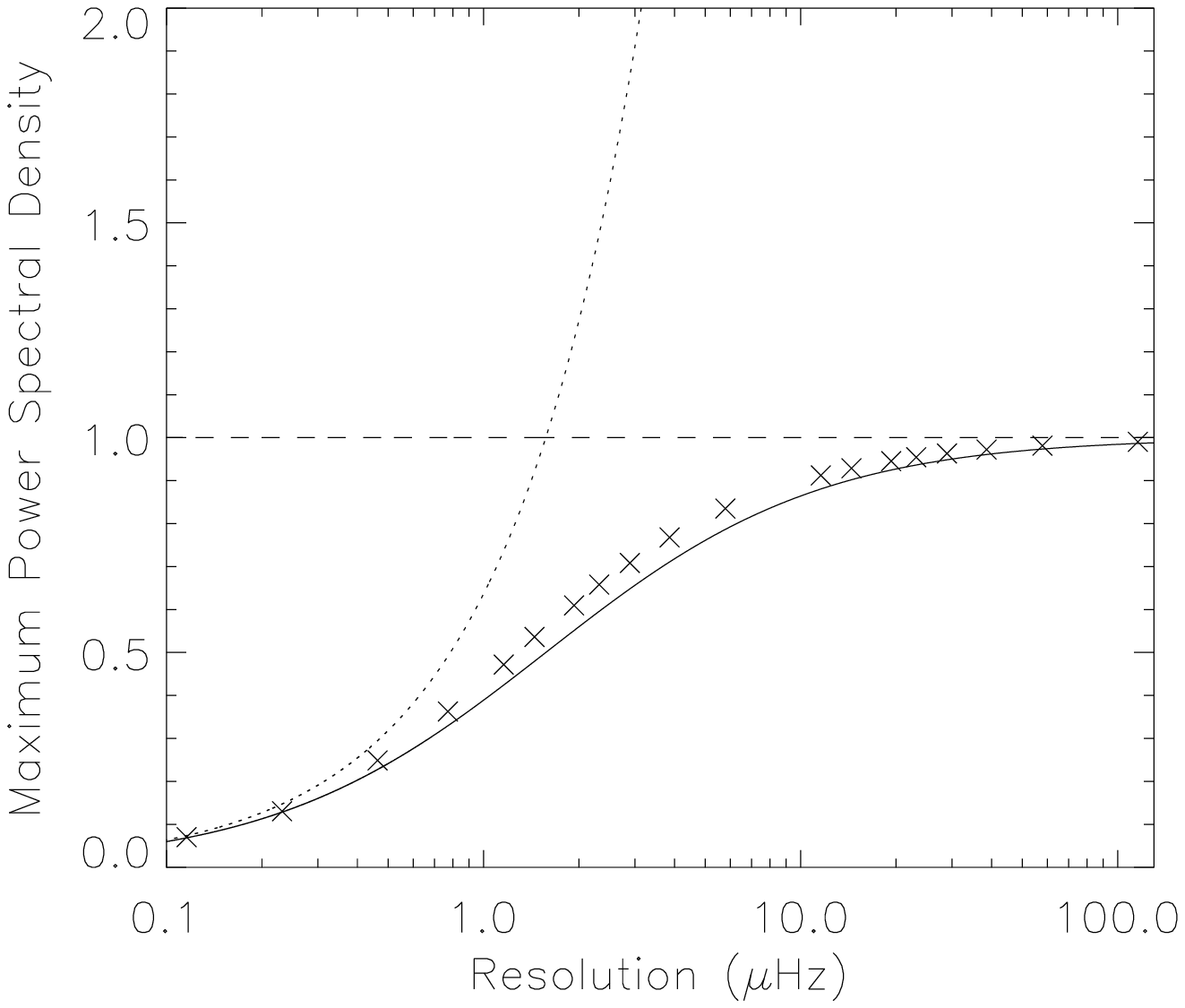}}\caption{Maximum power spectral density as a fraction of
the full modal power determined from the maximum value of 1000 co-added spectra (crosses). We also plot the
expected values as determined from the oversampled (dotted line); undersampled (dashed line) and modified (solid
line) expressions given in equations~\ref{HeightPowAll} and \ref{HeightPowMod}.} \label{HeightTest}
\end{figure}

However, were $T$ to be reduced to such an extent that $T \ll \tau$, the underlying profile would be so narrow
as to confine all power within a single bin. In this resulting undersampled regime we have conditions that tend
towards an undamped sine wave, where $H \sim P$. Therefore, a full description of $H$ in terms of the other
parameters is \citep{Chaplin2003}:
\begin{equation}
H = \left\{ \begin{array}{ll}
       2P(T/\tau) & \mbox{for $T \gg 2\tau$}\\
       P & \mbox{for $T \ll 2\tau$}.\end{array} \right.
       \label{HeightPowAll}
\end{equation}
Unfortunately, this does not give an adequate description for $H$ when $T \sim 2\tau$. In order to do this we
modify equation~\ref{HeightPowTau} slightly to give:
\begin{equation}
H =  \frac{2P}{(T/\tau)+2}, \label{HeightPowMod}
\end{equation}
which can be seen to work in both the over and undersampled regimes.

In order to test equation~\ref{HeightPowMod} we produced 1000 artificial time series of a single p-mode signal
with linewidth of 1 $\mu$Hz, ranging in length from 100 days down to 0.1 days. The power spectrum of each series
was then taken and the independent spectra co-added to produce a smooth peak allowing the maximum power spectral
density to be easily measured. In Fig.~\ref{HeightTest} we compare these values against those predicted from
equations~\ref{HeightPowAll} and \ref{HeightPowMod}. The figure shows clearly how the expressions given in
equation~\ref{HeightPowAll} diverge away from the true maximum power spectral density in the region $T \sim
2\tau$, whereas equation~\ref{HeightPowMod} matches reasonably closely throughout the entire range.

If one wishes to determine the power from the fitted linewidth, $\Delta$ and the maximum power spectral density
then equation~\ref{HeightPowMod} can be rearranged to give:
\begin{equation}
P = H \left(\frac{\pi}{2} T \Delta + 1 \right) \label{PowHeightMod1}
\end{equation}
which is the equation used in Section~\ref{PowSpec_Sec}. We add that even though the results presented in this
paper are in the range $T > 2\tau$, we still choose to use this more accurate expression.

\end{document}